\documentclass[aps,prd,preprintnumbers,showpacs,twocolumn,groupedaddress,nofootinbib]{revtex4}
\usepackage{graphicx}
\usepackage{latexsym}
\def\beq{\begin{equation}}
\def\eeq{\end{equation}}
\def\bey{\begin{eqnarray}}
\def\eey{\end{eqnarray}}

\def\lsim{\mathrel{\raise.3ex\hbox{$<$\kern-.75em\lower1ex\hbox{$\sim$}}}}
\def\gsim{\mathrel{\raise.3ex\hbox{$>$\kern-.75em\lower1ex\hbox{$\sim$}}}}

\begin{document}

\title{Dark Forces at the Tevatron}  
\author{Matt Buckley$^1$, Pavel Fileviez P\'{e}rez$^{2}$, Dan Hooper$^{1,3}$, and Ethan Neil$^{4}$}
\affiliation{$^1$Center for Particle Astrophysics, Fermi National Accelerator Laboratory, Batavia, IL 60510, USA}
\affiliation{$^2$Phenomenology Institute, Department of Physics, University of Wisconsin, 1150 University Avenue, Madison,  WI 53706, USA}
\affiliation{$^3$Department of Astronomy and Astrophysics, University of Chicago, Chicago, IL 60637, USA}
\affiliation{$^4$Particle Theory Group, Fermi National Accelerator Laboratory, Batavia, IL 60510, USA}

\date{\today}

\begin{abstract}

A simple explanation of the $W$+dijet excess recently reported by the CDF collaboration involves the introduction of a new gauge boson with sizable couplings to quarks, but with no or highly suppressed couplings to leptons. Anomaly-free theories which include such a leptophobic gauge boson must also include additional particle content, which may include a stable and otherwise viable candidate for dark matter. Based on the couplings and mass of the $Z'$ required to generate the CDF excess, we predict such a dark matter candidate to possess an elastic scattering cross section with nucleons on the order of $\sigma \sim 10^{-40}$ cm$^2$, providing a natural explanation for the signals reported by the CoGeNT and DAMA/LIBRA collaborations.  In this light, CDF may be observing the gauge boson responsible for the force which mediates the interactions between the dark and visible matter of our universe.

\end{abstract}

\pacs{95.35.+d, 14.70.Pw,14.80.-j; FERMILAB-PUB-11-180-A-T}
\maketitle

Very recently, the CDF collaboration announced the observation of an excess of events which include a lepton (electron or muon), missing transverse energy, and two jets. Within the context of the standard model, such events can arise from the production of a $W^{\pm}$ (which decays to a lepton and neutrino) along with an additional $Z$ or $W^{\pm}$ (which decays to a quark and an anti-quark, producing two jets). When the distribution of such events is plotted as a function of the invariant mass of the two jets, a sizable excess appears between 120 and 160 GeV. The significance of this excess (which is based on 4.3 fb$^{-1}$ of data) has been quoted as 3.2$\sigma$, including all known systematic uncertainties~\cite{cdf} (for more details pertaining to this analysis, see Ref.~\cite{cdfthesis}). 

Over the past several days, a number of possible explanations for the observed excess have been proposed~\cite{Buckley:2011vc,Yu:2011cw,Wang:2011uq,Cheung:2011zt,Eichten:2011sh,Kilic:2011sr,Wang:2011ta,Anchordoqui:2011ag,Nelson:2011us,Dobrescu:2011px,Sato:2011ui}. Most of these possibilities can be divided into two categories.  First, one could consider a scenario in which a new uncolored boson with a mass of 140-150 GeV is introduced, with couplings to light (and possibly other) quarks, but with little or no couplings to leptons --- a leptophobic $Z'$, for example~\cite{Buckley:2011vc,Yu:2011cw,Wang:2011uq,Cheung:2011zt,Wang:2011ta,Anchordoqui:2011ag,Nelson:2011us}. Second, there are models which make use of multiple new states. In these explanations of the CDF $W$+dijets anomaly, a particle is produced on resonance, which then decays to a $W^{\pm}$ along with a second new state with a mass of 140 to 150 GeV. Such a scenario can be realized, for example, within the context of technicolor~\cite{Eichten:2011sh} or R-parity violating supersymmetric models~\cite{Kilic:2011sr}. 

In this letter, we focus on the simplest of these possibilities -- a 140-150 GeV leptophobic $Z'$. A $Z'$ with suppressed couplings to leptons can appear in phenomenologically viable and well motivated extensions of the standard model. In particular, such a particle appears in models in which baryon number, instead of being a mere accidental global symmetry, is promoted to a local gauge symmetry~\cite{BL,BL2,Barger:1996kr}. In this case, the new vector boson couples to all standard model quarks with equal strength, but does not couple to leptons. In addition to providing a leptophobic $Z'$ as required by the CDF anomaly, this has the added benefit of suppressing proton decay~\cite{BL}. Other models which introduce a leptophobic $Z'$ rely on the kinetic mixing of two different $U(1)$'s to suppress the couplings to leptons. In this case, the $Z'$ bosons may have different couplings to up-type and down-type quarks, and even to quarks of different families~\cite{E6,Barger:1985dd,Barger:1996kr} (an attractive feature within the context of multi-$b$ excesses at the Tevatron~\cite{Buckley:2011vc}). Realizations of such a scenario can be found, for example, within the context of the $U(1)_{\eta}$ model of $E_6$ grand unification~\cite{E6}.

Any model containing new gauge bosons must also include new fermions (known as exotics) to cancel the corresponding anomalies (unless the couplings are proportional to those of the standard model gauge bosons). If baryon number and lepton number are each promoted to local gauge symmetries, the anomalies associated with $U(1)_B$ and $U(1)_L$ can each be cancelled by the introduction of one additional generation of fermions among which the new quarks (leptons) carry baryon number (lepton number) of magnitude 1 (3) (although the anomalies could instead be cancelled by three new generations with baryon and lepton number of 1/3 and 1, this leads to unacceptable flavor changing neutral currents and to tension with electroweak precision measurements)~\cite{BL,foot}. To avoid one or more of the exotic fermions being a cosmologically problematic heavy, stable and colored state, additional (non-fermionic) particle content must be added to mediate their decays. Such a particle must carry non-zero baryon number, and may be non-colored and electrically neutral. By virtue of baryon number conservation, such a state can be stable and a viable candidate for dark matter~\cite{BL}. In this letter, we explore the possible connection between the recent CDF anomaly and the dark matter of our universe, and discuss the associated phenomenology.

The precise coupling of this potential dark matter candidate (which we will from this point on denote as $X$) to the leptophobic $Z'$ is determined by its baryon number charge. While this is a somewhat model dependent quantity, one generally expect its baryon charge to be of order unity and thus the $Z'$ to couple to it and standard model quarks with similar strengths. In the model of Refs.~\cite{BL,BL2}, which we will discuss in more detail later, $X$ carries baryon number of 2/3, and thus couples to the $Z'$ with twice the strength of standard model quarks.

The couplings of the $Z'$ to quarks and the dark matter candidate, $X$, can lead to a potentially sizable spin-independent elastic scattering cross section between dark matter and nuclei. In particular, any real scalar or Dirac fermion dark matter particle interacting through a vector gauge boson will possess a cross section with nuclei given by:
\begin{eqnarray}
\sigma^{\rm SI}_{XN} = \frac{m^2_{X} m^2_N}{\pi (m_X+m_N)^2} \bigg[Z f_p + (A-Z)f_n \bigg]^2,  
\end{eqnarray}
where $A$ and $Z$ are the atomic mass and atomic number of the target nucleus and $f_{p,n}$ are the effective couplings to protons and neutrons:
\begin{equation}
f_p = \frac{g_{XXZ'} (2g_{uuZ'}+g_{ddZ'})}{m^2_{Z'}},\,\,\,\, f_n = \frac{g_{XXZ'} (g_{uuZ'}+2g_{ddZ'})}{m^2_{Z'}}. 
\end{equation}
Here, we have used $g_{XXZ'}$ and $g_{qqZ'}$ to denote the effective coupling strengths of the respective vertices. In the case in which the $Z'$ couples to baryon number, we can write these as $g_{qqZ'}=g_B/3$ and $g_{XXZ'}=Q_{B} \, g_B$, where $Q_B$ is the baryon number of the dark matter candidate and $g_B$ is the gauge coupling of the leptophobic $Z'$. 

Numerically, this leads to a spin-independent cross section with nucleons given by:
\begin{eqnarray}
\sigma^{\rm SI}_{Xp}&=&\sigma^{\rm SI}_{Xn}=\frac{m^2_{X} m^2_{p,n}}{\pi (m_X+m_{p,n})^2} \, \frac{g^4_B Q^2_B}{m^4_{Z'}}  \\
&\approx&  2 \times 10^{-40} \, {\rm cm}^2 \, \times \bigg(\frac{g_B}{0.3}\bigg)^4  \bigg(\frac{150 \, {\rm GeV}}{m_{Z'}}\bigg)^4  \bigg(\frac{Q_B}{1/3}\bigg)^2.\nonumber
\end{eqnarray}
Here, we have scaled to values of $m_{Z'}$ and $g_B$ which approximately lead to the observed properties of the CDF $W$+dijet excess~\cite{Buckley:2011vc}. For dark matter masses greater than $\sim10$ GeV, a cross section of this magnitude is in considerable conflict with constraints from CDMS~\cite{cdms}, XENON100~\cite{xenon100}, and other direct detection experiments (by more than three orders of magnitude for $m_{X}=100$ GeV, for example). For a lighter dark matter mass ($m_{X}\sim 5-10$ GeV), however, the constraints from direct detection experiments are much weaker~\cite{lightconstraints} (for a critical discussion, see Ref.~\cite{crit}). Furthermore, the CoGeNT~\cite{cogent} and DAMA/LIBRA~\cite{dama} collaborations report the observation of events which, although not attributable to any known backgrounds, could be explained by the elastic scattering of a dark matter particle with a mass of 5 to 10 GeV and a spin-independent cross section on the order of $\sigma^{\rm SI}_{Xp,Xn} \sim 10^{-40}$ cm$^2$~\cite{consistent}.

This has led us to the intriguing and suggestive result which we consider to be the main point of this letter: models which include a leptophobic gauge boson capable of accounting for the $W$+dijet excess recently reported by CDF must also invariably include additional matter, some of which is expected to couple to the $Z'$ with a similar strength as the $Z'$ couples to quarks. A stable and otherwise viable dark matter candidate can naturally appear among these states, and will likely possess an elastic scattering cross section with nuclei which is compatible with the signals reported by CoGeNT and DAMA/LIBRA.

Turning our attention now to the annihilation of dark matter particles, we find that we are unable to rely only on the exchange of the $Z'$ possibly observed by CDF. In particular, if the dark matter, $X$, is a Dirac fermion, then the $s$-channel exchange of the $Z'$ will lead to an annihilation cross section given by:
\begin{eqnarray}
\sigma v_{X\bar{X} \rightarrow Z' \rightarrow q\bar{q}} &\approx& \frac{5 g^4_{B} Q^2_B m^2_{X}}{3\pi m^4_{Z'}}  (1+v^2/6) \\
&\approx& 5.4\times 10^{-28} {\rm cm}^3/{\rm s} \nonumber \\
&\times& \bigg(\frac{g_B}{0.3}\bigg)^4  \bigg(\frac{150 \, {\rm GeV}}{m_{Z'}}\bigg)^4 \bigg(\frac{Q_B}{1/3}\bigg)^2 \bigg(\frac{m_X}{7 \, {\rm GeV}}\bigg)^2,\nonumber
\end{eqnarray}
which is well below the value of $\sigma v \approx 3\times 10^{-26}$ cm$^3$/s that is required to avoid the overproduction of dark matter in the early universe. If $X$ is a scalar, the exchange of a vector $Z'$ is velocity suppressed, making it even less effective at diluting the thermal abundance of dark matter. Annihilations through the standard model Higgs boson are also too small to provide an acceptable relic abundance~\cite{BL2}.

Other annihilation channels, however, could potentially be more efficient and lead to an acceptable thermal abundance of dark matter. Although such annihilation channels are quite model dependent, we will illustrate one example that appears in an existing model. 

In particular, we consider a model with a single additional generation ($e'_L$, $l'_R$, $Q'_R$, $u'_L$, $d'_L$, $\nu'_L$) with opposite chirality to the standard model fermions, and with baryon and lepton charges of 1 and 3, respectively (see model 2 of Refs.~\cite{BL,BL2}). In addition to an extended Higgs sector to break the baryon and lepton gauge symmetries, this model contains a scalar $S_L$ which is added to avoid lepton flavor violation, and a scalar $X$ which allows the heavy quarks to decay. $X$, which is electrically neutral, not colored, and carries baryon number $2/3$, is the dark matter candidate in this model. 

In this model, a pair of dark matter particles, $X$, can interact with two leptons according to:
\begin{equation}
G_{X e_i} X^\dagger X \bar{e}_i e_i,
\end{equation}
where
\begin{equation}
G_{X e_i}= \frac{\lambda_{X S_L}}{16 \pi^2} \left( \lambda_{l_i} \lambda_{e_i} +  \lambda_{l_i}^{\dagger} \lambda_{e_i}^{\dagger}  \right)  M_{e^{'}} C_0 (M_X, M_{S_L}, M_{e^{'}}).
\end{equation}
Here, $\lambda_{X S_L}$, $\lambda_{l_i}$, and $\lambda_{e_i}$ couplings which appear in
\begin{equation}
-{\cal L} \supset \lambda_{X S_L} X^{\dagger} X S_L^{\dagger} S_L + (\lambda_{e_i} \bar{e}_{R,i} S_L^{\dagger} e_L^{'}  +   \lambda_{l_i}  \bar{l}_R^{'} S_L l_{L,i}  +   \rm{h.c.}).
\end{equation} 
In order to evade constraints from $(g_{\mu}-2)$ and the branching fraction $\mu \rightarrow e \gamma$, we require that the above couplings to electrons and muons ($\lambda_{e_1}, \lambda_{e_2}, \lambda_{l_1}, \lambda_{l_2}$) be suppressed. For this reason, we will rely primarily on annihilations to taus (this will also enable us to evade constraints from LEP II, as described in Ref.~\cite{lepdm}).

The loop factor is given by
\begin{equation}
C_0 (M_X, M_{S_L}, M_{e^{'}}) = \int^1_0 dx \int^x_0  dy \ \chi (x,y)^{-1},  
\end{equation}
where
\begin{equation}
\chi (x,y)= -4 M_{X}^2 y (1-x) + M_{S_L}^2 y + M_{e^{'}}^2 (x-y) + M_{S_L}^2 (1-x).
\end{equation}
For masses of $M_X = 7$ GeV, $M_{S_L}=100$ GeV, and $M_{e^{'}}=200$ GeV one gets $C_0 \approx 2.8 \times 10^{-5}$. Taking $\lambda_{X S_L}, \lambda_{l_3}, \lambda_{e_3} \sim 1$ one gets $G_{X e_3} \approx 7.2 \times 10^{-5}$ GeV$^{-1}$. This interaction leads to a contribution to the dark matter's annihilation cross section given by:
\begin{eqnarray}
\sigma v_{XX\rightarrow l^+ l^-} &=& \frac{1}{4 \pi} \sum_{i=1}^3 |G_{X e_i}|^2 \\ \nonumber
&\sim& 3 \times 10^{-26} \rm{cm}^{3}/{\rm s} \, \times \bigg(\frac{\lambda_{e_3}}{1.5}\bigg)^2  \, \bigg(\frac{\lambda_{l_3}}{1.5}\bigg)^2,
\end{eqnarray}
which is consistent to the value required of a thermal relic. Although the annihilation channels which may be available to a dark matter candidate in a model containing a leptophobic $Z'$ are somewhat model dependent, this example illustrates that sufficiently efficient channels can naturally appear in such models.

Annihilation processes such as that described above not only enable the dark matter to be efficiently annihilated in the early universe, but can also provide potentially observable fluxes of dark matter annihilation products. In the example described above, the dark matter will annihilate in large part to charged leptons. Although one might naively expect dark matter to have large annihilation cross sections to baryons in models with a leptophobic $Z'$, this is not the case (barring gauge couplings much larger than those required to generate the CDF $W$+dijets anomaly). As demonstrated above, however, in leptophobic $Z'$ models in which baryon and/or lepton charges are gauged, it is natural for dark matter to annihilate primarily to leptons. 

This feature is particularly attractive within the context of observations of the Galactic Center by the Fermi Gamma Ray Space Telescope, which are consistent with being the product of dark matter annihilations which proceed primarily to $\tau^+ \tau^-$, among other leptonic final states~\cite{lisa}. Similarly, 5-10 GeV dark matter particles which annihilate primarily to leptons are predicted to generate synchrotron emission from the Inner Milky Way consistent with that observed by WMAP~\cite{hazelight,haze}.
 
In the near future, a number of experiments will provide ways to test aspects of the scenario described in this letter. First of all, if a $Z'$ is responsible for the CDF anomaly, signals should also appear (although likely with lesser statistical significance) in channels including $(Z\rightarrow l^+l^-)+(Z'\rightarrow jj)$, $(Z\rightarrow \nu \bar{\nu})+(Z'\rightarrow jj)$, and $\gamma+(Z'\rightarrow jj)$. Furthermore, the invisible decays of the $Z'$ to dark matter will also lead to events of the type $(Z\rightarrow l^+l^-)+(Z'\rightarrow XX)$. The rate of such events will be suppressed by the branching fraction of the $Z'$ to dark matter, however, which we estimate to be $1.6\% \,\times \, (Q_B/0.33)^2$ if the dark matter is a scalar and $6.3\% \,\times \, (Q_B/0.33)^2$ if it is a Dirac fermion. Additionally, if the events reported by CoGeNT are the result of an elastically scattering dark matter particle, their existing 15 months of data should be sufficient to identify (or rule out) a statistically significant annual modulation~\cite{consistent,cogentmod}.

In summary, we have considered in this letter a leptophobic $Z'$ as an explanation of the $W$+dijets excess reported by CDF, and discussed the possible connection of this particle to the dark matter of our universe. In particular, the introduction of a $Z'$ with couplings to standard model quarks (but not leptons) must be accompanied by new matter which assists in the cancellation of anomalies. Among these new states may naturally appear a candidate for dark matter (for a well-defined example of such a model, see Refs.~\cite{BL,BL2}). If the couplings of the $Z'$ to dark matter are similar to its couplings to quarks, one predicts an elastic scattering cross section between dark matter and nucleons on the order of $\sigma \sim 10^{-40}$ cm$^2$, providing a explanation for the signals reported by the CoGeNT and DAMA/LIBRA collaborations.  
Although annihilations of dark matter through the exchange of the $Z'$ in this scenario does not yield a cross section large enough to provide an acceptable thermal relic abundance, other annihilation channels can potentially be much more efficient. While the details of this process are model dependent, we have described an example of a viable annihilation process which appears within the context of the models described in Ref.~\cite{BL,BL2}. In this particular case, we find that dark matter annihilates primarily to tau leptons, and thus could account for the gamma ray signal observed from the Inner Galaxy by the Fermi Gamma Ray Space Telescope.

\smallskip

We would like to thank Mark Wise, Joachim Kopp, and Graham Kribs for valuable discussions. PFP would like to thank the Center for the Fundamental Laws of Nature at Harvard University and the Center for Theoretical Physics at MIT for their hospitality during the completion of this article. MRB and DH are supported by the US Department of Energy and by NASA grant NAG5-10842. PFP is supported in part by the US Department of Energy under contract DE-FG02-95ER40896.

\end{document}